\documentclass[twoside]{elsart}
\usepackage{epsfig}
\usepackage{array}
\usepackage{amssymb,amsbsy,amsmath}
\usepackage{times}
\newcommand{\fmref}[1]{(\protect\ref{#1})}
\def\half{{\frac{1}{2}}\,}
\newcommand{\SmallMean}[1]{\langle\langle \; {#1}\; 
            \rangle\rangle}
\begin{document}
%
\typeout{   --- >>>   2. ho paper   <<<   ---   }
\typeout{   --- >>>   2. ho paper   <<<   ---   }
\typeout{   --- >>>   2. ho paper   <<<   ---   }
%
%
\journal{PHYSICA A}
\begin{frontmatter}
\title{Thermodynamic fermion-boson symmetry in harmonic oscillator potentials}
 
\author{H.-J. Schmidt and J. Schnack\thanksref{JS}}
\address{Universit\"at Osnabr\"uck, Fachbereich Physik \\ 
         Barbarastr. 7, D-49069 Osnabr\"uck}

\thanks[JS]{hschmidt\char'100uos.de, jschnack\char'100uos.de,\\
\phantom{xx}http://www.physik.uni-osnabrueck.de/makrosysteme/}

\begin{abstract}
\noindent
A remarkable thermodynamic fermion-boson symmetry is found for
the canonical ensemble of ideal quantum gases in harmonic
oscillator potentials of odd dimensions. The bosonic partition
function is related to the fermionic one extended to negative
temperatures, and vice versa.

\vspace{1ex}

\noindent{\it PACS:} 
05.30.-d; 
05.30.Ch; 
05.30.Fk; 
05.30.Jp 

\vspace{1ex}

\noindent{\it Keywords:} Quantum statistics; Canonical ensemble;
Finite Fermi and Bose systems
\end{abstract}
\end{frontmatter}
\raggedbottom
\section{Introduction and summary}

Finite quantum gases of identical fermions or bosons contained
in harmonic oscillator potentials are currently of great
interest. The recent motivation stems from the investigation of
alkali atoms (bosons) in magnetic traps and the observation of
Bose-Einstein condensation, see
e.g. \cite{AEM95,DMA95,BSH97}. On the fermionic side nuclei and
their low temperature behaviour are investigated, for instance
in \cite{Poc95,ScF97}. If these systems are either only weekly
interacting or close to the ground state, they may be considered
as an ideal gas of (quasi-) particles confined by a harmonic
oscillator potential.

For these finite ideal quantum gases simple expressions,
e.g. for the partition function of the canonical ensemble, could
not be found up to now, but recursion relations could be derived
\cite{BoF93} and used successfully
\cite{BLD96,BLD98,GrH97,WiW97,ScS98}. 

In this article we report that a closer inspection of the
canonical partition function uncovers a surprising symmetry
property which connects fermions and bosons. Fermionic and
bosonic $N$-body partition functions are connected via analytic
continuation to negative temperatures. Thermodynamic mean values
for the bosonic systems can be easily evaluated knowing the
fermionic results and vice versa. Especially for fermions this
procedure removes the well-known sign problem because the
bosonic partition function does not suffer from alternating
signs.

\section{Symmetries of the partition function}

We consider $N$ identical particles subject to a single-particle
Hamiltonian in the canonical ensemble.  The partition function
$Z_N$ can be recursively built starting with the single-particle
partition function \cite{BoF93}
\begin{eqnarray}
\label{E-2-1}
Z_N^{\pm}(\beta)
=
\frac{1}{N} \sum_{n=1}^N \; (\pm 1)^{n+1}\;
Z_1(n \beta) \; Z_{N-n}^{\pm}(\beta)
\ , \quad
Z_0(\beta) = 1
\ , \
\beta = \frac{1}{k_B T}
\ ,
\end{eqnarray}
where the upper sign in the sum stands for bosons and the lower
sign for fermions. For the case of a three-dimensional harmonic
oscillator potential the single-particle partition function is 
\begin{eqnarray}
\label{E-2-2}
Z_1(\beta) 
=
\left[
\frac{\exp\left(-\beta \frac{\hbar\omega}{2} \right)}
     {1 - \exp(-\beta \hbar\omega)} 
\right]^3
\ .
\end{eqnarray}
Introducing the abbreviation
\begin{eqnarray}
\label{E-2-3}
y = \exp(-\beta \hbar\omega)
\end{eqnarray}
we redefine the partition functions to depend on $y$, e.g.
\begin{eqnarray}
\label{E-2-4}
Z_1(y) 
=
\frac{y^{\frac{3}{2}}}{(1 - y)^3}
\ .
\end{eqnarray}
The recursion formula \fmref{E-2-1} then reads
\begin{eqnarray}
\label{E-2-5}
Z_N^{\pm}(y)
=
\frac{1}{N} \sum_{n=1}^N \; (\pm 1)^{n+1}\;
Z_1(y^n) \; Z_{N-n}^{\pm}(y)
\ .
\end{eqnarray}
The partition functions $Z_N^{\pm}$ are rational functions of
$y$ and can be split into factors \cite{BLD96,BLD98}
\begin{eqnarray}
\label{E-2-6}
Z_N^{\pm}(y)
=
\frac{y^{\frac{3N}{2}}}{\prod_{j=1}^{N}(1-y^j)^3}\;
P_N^{\pm}(y)
\, 
\end{eqnarray}
with $P_N^{\pm}(y)$ being polynomials in $y$.  Obviously this
leads to a recursion formula for the polynomials $P_N^{\pm}$
\begin{eqnarray}
\label{E-2-7}
P_N^{\pm}(y)
=
\frac{1}{N} \sum_{n=1}^N \; (\pm 1)^{n+1}\;
\frac{\prod_{j=N-n+1}^{N}(1-y^j)^3}{(1-y^n)^3}\;
P_{N-n}^{\pm}(y)
\ ,
\end{eqnarray}
where $P_0^{\pm}(y)=1$ and $P_1^{\pm}(y)=1$. This is already a
useful result since for numerical evaluation recursion
formula \fmref{E-2-7} behaves better.

A closer look at these polynomials opens new insight into
fermion-boson symmetry. Consider for example $N=3$:
\begin{eqnarray}
\label{E-2-8}
P_3^{+}(y)
&=&
1 + 3 y^2 + 7 y^3 + 6 y^4 + 6 y^5 + 10 y^6 + 3 y^7
\\
P_3^{-}(y)
&=&
y^2 \left(
3 + 10 y + 6 y^2 + 6 y^3 + 7 y^4 + 3 y^5 + 1 y^7
\right)
\nonumber
\ .
\end{eqnarray}
It seems surprising that the coefficients appear in reverse
order comparing one polynomial with the other. This is not mere
coincidence, but the following theorem holds
\begin{eqnarray}
\label{E-2-9}
P_N^{-}(y)
=
y^{\frac{3}{2}N(N-1)}\;
P_N^{+}\left(\frac{1}{y}\right)
\ ,
\end{eqnarray}
which can be proven by induction or with the help of an explicit
representation as done below. The property of the bosonic and
fermionic polynomials leads to a relation between fermionic and
bosonic partition functions which is
\begin{eqnarray}
\label{E-2-10}
Z_N^{+}(y)
=
(-1)^N
Z_N^{-}\left(\frac{1}{y}\right)
\ .
\end{eqnarray}
Using the usual dependence on the inverse temperature $\beta$
this property expresses itself as
\begin{eqnarray}
\label{E-2-11}
Z_N^{+}(\beta)
=
(-1)^N
Z_N^{-}(-\beta)
\ ,
\end{eqnarray}
where the partition function with the negative argument has to
be understood as the analytic continuation into the region of
$y=\exp(-\beta \hbar\omega)>1$.  In thermodynamic mean values
like mean energy or specific heat this symmetry also shows up
\begin{eqnarray}
\label{E-2-12}
E_N^{+}(\beta)
=
- E_N^{-}(-\beta)
\quad ,
\qquad
C_N^{+}(\beta)
=
C_N^{-}(-\beta)
\ .
\end{eqnarray}
A straight forward application of the above result is to
calculate fermionic partition functions and mean values by
evaluating the respective bosonic ones at negative temperatures
and thereby to avoid the sign problem.

More generally this property is related to the fact that the
single-particle partition function has an analytic continuation
to the whole $\beta$-axis where it is an odd function,
$Z_1(-\beta)=-Z_1(\beta)$. Then $Z_N^{\pm}$ can be analytically
extended in the same way and it satisfies eq. \fmref{E-2-11}.  We
prove this relation rewriting the explicit representation of
$Z_N^{\pm}(\beta)$ as it is given in eq.~(8) of
ref. \cite{ScS98} in the form
\begin{eqnarray}
\label{E-2-13}
Z_N^{\pm}
=
(\pm 1)^N
\sum_{k=0}^{N-1}
\sum_{0 < n_1 < \cdots < n_k < N}
(\pm 1)^{k+1}
\frac{\prod_{i=0}^{k}Z_1\left( (n_{i+1}-n_i) \beta\right)}
     {\prod_{j=1}^{k} n_j}
\ ,
\end{eqnarray}
where we understand $n_0=0$ and $n_{k+1}=N$ in the upper product.
If $\beta$ is replaced by $-\beta$ we obtain $(k+1)$ minus signs
in the upper product, which either cancel the factor
$(-1)^{k+1}$ in the fermionic case or introduce it in the
bosonic one and thereby transform the fermionic partition
function into the bosonic one or vice versa.

This theorem shows that the fermion-boson symmetry depends only
on the oddness of $Z_1$ and not on the form of the
single-particle Hamiltonian. But the harmonic oscillator potentials
for odd space dimensions are the only examples we know, where
this condition holds.

\section{Properties of the polynomials}

In order to study the thermodynamics of ideal quantum gases in
harmonic oscillator potentials one would like to know more about
the polynomials $P_N^{\pm}$ defined by \fmref{E-2-6}. Using the
explicit representation of the partition function
\fmref{E-2-13} one derives 
\begin{eqnarray}
\label{E-3-1}
P_N^{\pm}(y)
&=&
(\pm 1)^N
\prod_{l=1}^{N} (1-y^l)^3\;
\\
&&\times
\sum_{k=0}^{N-1}
\sum_{0 < n_1 < \cdots < n_k < N}
(\pm 1)^{k+1}
\frac{1}
     {\prod_{j=1}^{k} n_j \; \prod_{i=0}^{k}\left( 1 -
     y^{n_{i+1}-n_i}\right)^3}
\nonumber
\ .
\end{eqnarray}
Note, that $P_N^{\pm}$, although looking like a rational
function, is actually a polynomial since
$\prod_{i=0}^{k}\left(1-y^{n_{i+1}-n_i}\right)^3$ divides 
$\prod_{l=1}^{N}(1-y^l)^3$.

In the following we have to distinguish between properties of
$P_N^{\pm}$, which can be mathematically proven and those which
are only established by numerical evidence, i.e. usually checked
with MATHEMATICA for several $N$. For simplicity we are
considering only bosonic polynomials now, the respective
fermionic ones can be obtained by transformation.

The following belongs to the second class (numerical
evidence). The coefficients $p_n^{(N)}$ of the polynomials
\begin{eqnarray}
\label{E-3-2}
P_N^{+}(y)
&=&
\sum_{n=0}^{L(N)}
p_n^{(N)}\; y^n
\end{eqnarray}
are non-negative and, for $n>0$ and $N>4$, monotonically
increasing until a maximum is reached and then monotonically
decreasing. The degree $L(N)$ satisfies 
\begin{eqnarray}
\label{E-3-3}
L(N)
\le
\half (3 N^2 -7 N + 10)
\ .
\end{eqnarray}
The first $N+1$ coefficients $p_n^{(N)}$ are independent of
$N$. They form an approximately exponentially increasing sequence
\begin{eqnarray}
\label{E-3-4}
1, 0, 3, 7, 18, 39, 99, 213, 492, 1056, \dots
\ .
\end{eqnarray}
For the independence of the first $N/2$ coefficients we know a
mathematically exact proof, for the others we have numerical
evidence. 

Several identities hold for the polynomials, e.g.
\begin{eqnarray}
\label{E-3-5}
P_N(1) &=& (N!)^2
\\
P_N^{\prime}(1) &=& \frac{3}{4} N (N-1) (N!)^2
\nonumber \\
P_N^{\prime\prime}(1) &=& \frac{1}{48} (27 N^2 - 23 N - 8) 
N (N-1) (N!)^2
\nonumber \\
P_N^{\prime\prime\prime}(1) &=& \frac{1}{64} (27 N^4 - 42 N^3 - 9
N^2 +16 N -20) N (N-1) (N!)^2
\nonumber
\ ,
\end{eqnarray}
where the primes denote derivatives. We sketch a proof for the
first relation,
\begin{eqnarray}
\label{E-3-6}
P_N(1) 
=
\sum_{n=0}^{L(N)}
p_n^{(N)}
= 
(N!)^2
\ ,
\end{eqnarray}
the others may be proven in a similar way.
All terms with $n>1$ in recursion formula \fmref{E-2-7} contain
at least one factor $(1-y^j)^3$ and thus vanish for $y=1$. Hence
\begin{eqnarray}
\label{E-3-7}
P_N(1) 
&=&
\frac{1}{N}
\frac{(1-y^N)^3}{(1-y)^3} P_{N-1}(y) \; {\big |}_{y=1}
\\
&=&
\frac{1}{N}
(1+y+y^2+\cdots +y^{N-1})^3 P_{N-1}(y) \; {\big |}_{y=1}
\nonumber \\
&=&
N^2 P_{N-1}(1)
\nonumber
\end{eqnarray}
The relation \fmref{E-3-6} then follows by $P_1(1)=1$ and
induction.

Introducing a quasi-statistical notation
\begin{eqnarray}
\label{E-3-8}
\SmallMean{f(n)}
:=
\frac{1}{(N!)^2}
\sum_{n=0}^{L(N)}
p_n^{(N)}
f(n)
\end{eqnarray}
we can summarize these results as
\begin{eqnarray}
\label{E-3-9}
\mu 
&:=&
\SmallMean{n}
=
\frac{3}{4} N (N-1)
\\
\sigma^2
&:=&
\SmallMean{n^2} - \mu^2
=
\frac{1}{24} (2 N + 5) N (N-1)
\nonumber \\
\mu_3
&:=&
\SmallMean{(n-\mu)^3}
=
-\frac{3}{8} N (N-1)
\nonumber
\ .
\end{eqnarray}
The last result is remarkable since it implies that the
``skewness" of the distribution of the $p_n^{(N)}$, defined by
\cite{Abr}
\begin{eqnarray}
\label{E-3-10}
\gamma 
=
\frac{\mu_3}{\sigma^3}
\end{eqnarray}
is asymptotically proportional to $N^{-5/2}$ and thus vanishes
for $N\rightarrow\infty$. This is compatible with the
observation that for large $N$ the distribution of coefficients
can be approximated by a symmetric bell-shaped function.


{\bf Acknowledgments}\\[5mm]
We thank Klaus B\"arwinkel for carefully reading
the manuscript.

\end{document}